\documentclass[aps,prl,twocolumn,superscriptaddress]{revtex4-1}

\usepackage{graphicx}
\usepackage{amsmath}
\usepackage{amssymb}
\usepackage{epstopdf}
\usepackage{color}
\usepackage{lipsum}
\usepackage{hyperref}
\usepackage{ulem}
\usepackage[dvipsnames]{xcolor}

\DeclareGraphicsExtensions{.eps}

\newcommand*{\rom}[1]{\expandafter\@slowromancap\romannumeral #1@}

\begin{document}


\title{Interfacing scalable photonic platforms: solid-state based multi-photon interference in a reconfigurable glass chip}



\author{C. Ant\'on}
\email[]{carlos.anton-solanas@c2n.upsaclay.fr}
\affiliation{Center of Nanosciences and Nanotechnology - CNRS, Univ. Paris-Saclay, UMR 9001, 10 Boulevard Thomas Gobert, 91120 Palaiseau France}

\author{J. C. Loredo}
\affiliation{Center of Nanosciences and Nanotechnology - CNRS, Univ. Paris-Saclay, UMR 9001, 10 Boulevard Thomas Gobert, 91120 Palaiseau France}

\author{G. Coppola}
\affiliation{Center of Nanosciences and Nanotechnology - CNRS, Univ. Paris-Saclay, UMR 9001, 10 Boulevard Thomas Gobert, 91120 Palaiseau France}

\author{H. Ollivier}
\affiliation{Center of Nanosciences and Nanotechnology - CNRS, Univ. Paris-Saclay, UMR 9001, 10 Boulevard Thomas Gobert, 91120 Palaiseau France}

\author{N. Viggianiello}
\affiliation{Dipartimento di Fisica, Sapienza Universit\`a di Roma, Piazzale Aldo Moro 5, Roma I-00185, Italy}

\author{A. Harouri}
\affiliation{Center of Nanosciences and Nanotechnology - CNRS, Univ. Paris-Saclay, UMR 9001, 10 Boulevard Thomas Gobert, 91120 Palaiseau France}

\author{N. Somaschi}
\affiliation{Quandela, SAS, 86 Rue de Paris, 91400 Orsay, France}

\author{A. Crespi}
\affiliation{Istituto di Fotonica e Nanotecnologie, CNR, P. Leonardo da Vinci, 32, Milano I-20133, Italy}
\affiliation{Dipartimento di Fisica, Politecnico di Milano, Piazza Leonardo da Vinci, 32, Milano I-20133, Italy}

\author{I. Sagnes}
\affiliation{Center of Nanosciences and Nanotechnology - CNRS, Univ. Paris-Saclay, UMR 9001, 10 Boulevard Thomas Gobert, 91120 Palaiseau France}

\author{A. Lema\^ itre}
\affiliation{Center of Nanosciences and Nanotechnology - CNRS, Univ. Paris-Saclay, UMR 9001, 10 Boulevard Thomas Gobert, 91120 Palaiseau France}

\author{L. Lanco}
\affiliation{Center of Nanosciences and Nanotechnology - CNRS, Univ. Paris-Saclay, UMR 9001, 10 Boulevard Thomas Gobert, 91120 Palaiseau France}
\affiliation{Universit\'e Paris Diderot - Paris 7, 75205 Paris CEDEX 13, France}

\author{R. Osellame}
\email[]{roberto.osellame@polimi.it}
\affiliation{Istituto di Fotonica e Nanotecnologie, CNR, P. Leonardo da Vinci, 32, Milano I-20133, Italy}
\affiliation{Dipartimento di Fisica, Politecnico di Milano, Piazza Leonardo da Vinci, 32, Milano I-20133, Italy}

\author{F. Sciarrino}
\email[]{fabio.sciarrino@uniroma1.it}
\affiliation{Dipartimento di Fisica, Sapienza Universit\`a di Roma, Piazzale Aldo Moro 5, Roma I-00185, Italy}

\author{P. Senellart}
\email[]{pascale.senellart-mardon@c2n.upsaclay.fr}
\affiliation{Center of Nanosciences and Nanotechnology - CNRS, Univ. Paris-Saclay, UMR 9001, 10 Boulevard Thomas Gobert, 91120 Palaiseau France}

\date{\today}

\begin{abstract}
{Scaling-up optical quantum technologies requires to combine highly efficient multi-photon sources and integrated waveguide components. Here, we interface these scalable platforms: a quantum dot based multi-photon source and a reconfigurable photonic chip on glass are combined to demonstrate high-rate three-photon interference. The temporal train of single-photons obtained from a quantum emitter is actively demultiplexed to generate a $3.8$~kHz three-photon source, which is then sent to the input of a tuneable tritter circuit, demonstrating the on-chip quantum interference of three indistinguishable single-photons. Pseudo number-resolving photon detection characterising the output distribution shows that this first combination of scalable sources and reconfigurable photonic circuits compares favourably in performance with respect to previous implementations. A detailed loss-budget shows that merging solid-state based multi-photon sources and reconfigurable photonic chips could allow ten-photon experiments on chip  at ${\sim}40$~Hz rate in a foreseeable future.}
\end{abstract}


\maketitle

\section{\label{sec:intro} Introduction}

The development of optical quantum technologies allows for quantum-enhanced metrology, secure quantum communication, and quantum computing and simulation~\cite{OBrien:2009aa,Aspuru-Guzik:2012aa,acin_quantum_2018} in highly-increased dimensions. Maturing quantum photonics requires efficient generation and detection of single-photons, as well as their scalable manipulation~ \cite{flamini_photonic_2019}. Single-photon detection is a well-advanced technology to date, and has already reached near-optimal values in efficiencies~\cite{Marsili:2013aa}. For single-photon generation, significant advances have been demonstrated using heralded approaches based on frequency conversion~\cite{zhong_12-photon_2018,spring_chip-based_2017}; however, their single-photon purity unavoidably decreases with the source brightness, which is defined as the probability $p_1$ of providing a single-photon per excitation pulse~\cite{Senellart:2017aa}. Temporal multiplexing schemes have been explored to circumvent this limitation~\cite{kaneda_high-efficiency_2018,kaneda_time-multiplexed_2015}, but at the expense of overall operation rates. On the other hand, spatial multiplexing schemes of many heralded sources~\cite{zhong_12-photon_2018,spring_chip-based_2017} implies a dramatic increase of resource overhead. 

Recently, scalable technologies for single-photon generation have emerged using quantum dots (QDs) in microcavities~\cite{somaschi_near-optimal_2016,ding_-demand_2016,Senellart:2017aa,liu_solid-state_2019}, where a single artificial atom emits temporal trains of single photons on demand. The  brightness already exceeds by more than one order of magnitude that of heralded sources of the same quality and near-deterministic sources could be reached with a similar technology and modified excitation scheme~\cite{he_polarized_2018}. This new generation of sources has allowed multiphoton experiments such as Boson Sampling~\cite{loredo_boson_2017,wang_high-efficiency_2017} involving up to five detected single-photons \cite{Wang:2018ab}.

{Photon manipulation can suffer from mechanical instabilities in bulk circuits, which lead to optical phase drifts and induces errors in device performance. A scalable photonic platform should instead provide photon routing and control in low-loss, integrated, and reconfigurable chips. These devices have been developed using various materials, such as silicon~\cite{gates_unconventional_2004,doi:10.1021/cr030076o,carolan_universal_2015}, silicon nitride~\cite{roeloffzen_silicon_2013}, lithium niobate~\cite{xiong_integrated_2011,jin_-chip_2014}, or  glass~\cite{valle_micromachining_2009,MALINAUSKAS20131,sugioka_femtosecond_2014}. The latter, based on femtosecond laser writing, offers fast and cheap production, and has been used to tackle a variety of complex quantum operations, such as Boson Sampling~\cite{crespi_integrated_2013,spagnolo_experimental_2014,bentivegna_experimental_2015}, quantum Fourier transforms~\cite{crespi_suppression_2016}, and quantum walks~\cite{sansoni_two-particle_2012,crespi_anderson_2013}}. In addition, this technique has shown great versatility in terms of polarization control \cite{corrielli_rotated_2014} and 3D patterning \cite{crespi_suppression_2016}; thermally tuneable phase-shifters can also be conveniently integrated to achieve circuit reconfigurability \cite{flamini_thermally_2015, crespi_single-photon_2017}. Despite spectacular progress on both solid-state photon sources, and reconfigurable photonic circuits on chip, these two promising platforms have not yet been combined---an approach that can result crucial for scaling optical quantum technologies.

\begin{figure*}[hbt!]
\centering
\includegraphics[width=\textwidth]{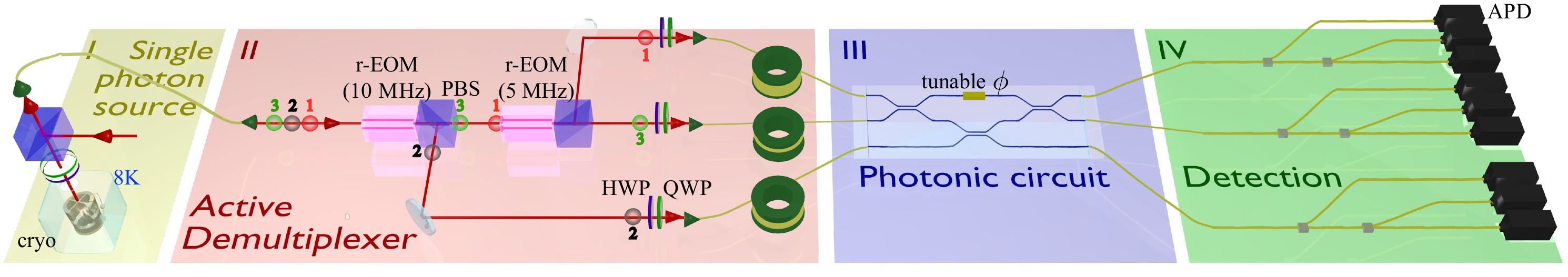}\vspace{-3mm}	
\caption{\textbf{Experimental scheme of the efficient three photon coalescence.} (I) Generation of single-photons from a QD-micropillar device under resonant fluorescence excitation. (II) Preparation of the three photons simultaneously arriving to the tritter input via active demultiplexing. (III) Circuit of the tritter providing the three photon coalescence. (IV) Detection of the quantum state of light at the output of the tritter.}
\label{fig:1}
\end{figure*}

In this work, we interconnect both scalable photonic platforms: we observe three-photon coalescence using an efficient solid-state based multi-photon source, and a reconfigurable photonic tritter circuit on glass. 
This first implementation of joint platforms already shows an improved performance in terms of quantity and quality of the three-photon interference. Furthermore, we estimate that feasible improvements can allow for ten-photon experiments at ${\sim}40$~Hz rates in the near future.

The experimental scheme, sketched in Fig.~\ref{fig:1}, is composed by four modules: (I) \textbf{single-photon generation} at  high rates  from a  QD source, (II) \textbf{time-to-spatial active demultiplexing} to prepare a three-photon source, (III) \textbf{photonic circuit} of the reconfigurable tritter, and (IV) \textbf{detection} of the photonic state via pseudo number-resolving measurements. In the following, we present each module before discussing the performances of the combined system.

\section{\label{sec:gen} Single-photon source}

We use a solid-state single-photon source consisting of a single InGaAs/GaAs QD deterministically coupled, with nanometric accuracy, to a micropillar cavity \cite{dousse_controlled_2008}. The micropillar device is gradually doped in the vertical direction to form a p-i-n diode structure embedding the QD in the intrinsic region of the cavity. Electrical contacts are defined in the top and bottom parts of the pillar to gain tunability of the QD energy via the confined Stark effect~\cite{nowak_deterministic_2014,somaschi_near-optimal_2016}. The device is mounted in a closed-cycle cryostat at ${\sim}8$ K, and an optical confocal cross-polarisation setup is used to excite the single-photon source with a resonant pulsed laser, see Fig.~\ref{fig:1}~(I). The experiments are performed with a neutral exciton in resonance with the cavity mode, which presents a single-photon lifetime of 160 ps and a wavelength of 925.47 nm. This short single-photon lifetime allows increasing  the laser repetition rate (81 MHz) by a factor of 4 using a passive pulse multiplier composed of BSs and delay lines~\cite{Broome:11}. 

The first lens brightness of  the source is measured to be $p_1{=}16.0{\%}$ in line with state of the art performances for neutral exciton~\cite{Senellart:2017aa}. The fibered brightness is measured to be  $p_1{=}7.0{\%}$, where this value is limited by a finite numerical aperture N.A.{=}0.45 of the first lens. This corresponds to a generation rate of $22$ MHz of single-photons in a single-mode fiber for a 324~MHz pumping rate. Standard photon correlation measurements are used to characterise the source performance under a repetition rate of 81~MHz. The single-photon purity is found to be $g^{(2)}(0){=}0.035{\pm}0.003$, and the photon indistinguishability corrected (uncorrected) from $g^{(2)}(0)$ is measured to be $M{=}0.920{\pm}0.007$ ($0.850{\pm}0.007$) for ${\sim}12.5$~ns between emitted photons, and $0.880{\pm}0.009$ ($0.810{\pm}0.009$) for ${\sim}100$~ns maximum temporal distance between the emitted photons subsequently used for interference. 

\section{\label{sec:prep} Demultiplexer}

\begin{figure*}[hbt!]
\centering
\includegraphics[width=1\textwidth]{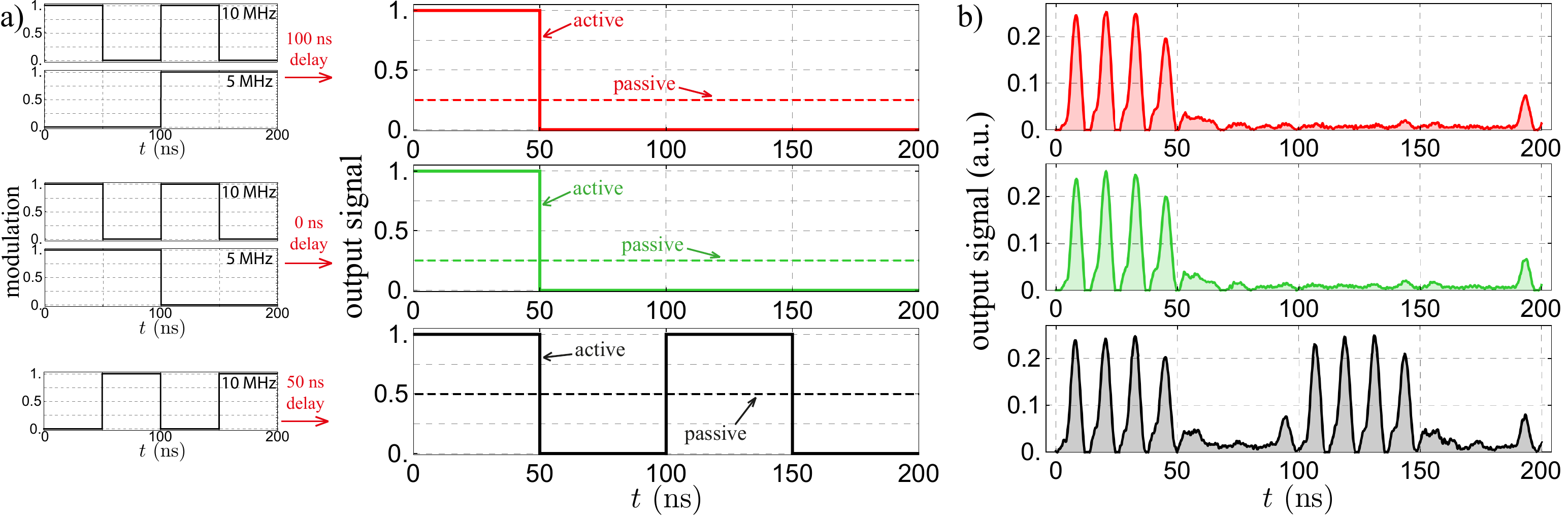}\vspace{-3mm}	
\caption{\textbf{Operation of the demultiplexer.} (a) The interplay of 2 modulators, one run at 10~MHz, and the other at 5~MHz, result in the relative output signals (red, green,black) after suitable delays. Dashed lines depict the same signals when the modulator is turned off (passive). (b) Measured output signals, resembling a similar behaviour to that presented in panel (a).}
\label{fig:2}
\end{figure*}

Multi-photon interference requires the efficient preparation of indistinguishable single-photons arriving simultaneously to the input ports of a photonic circuit~\cite{Latmiral:2016aa,wang_high-efficiency_2017,Wang:2018ab}. One can build the required multi-photon sources starting from a single-photon source via a demultiplexer: a device that routes its input train of temporal modes into separate and simultaneous spatial modes. In general, a demultiplexer can be characterised by their relative time-varying output signals. The device is a passive demultiplexer if these signals are static in time, and its conversion rate---the ratio between the output $n$-photon event rates, and the input single-photon rate---is given by $\mathcal{C}_n^\text{(passive)}{=}\prod_{k=1}^{n}p^\text{out}_k$, where $p^\text{out}_k$ is the static probability of the input signal to exit the $k{-}th$ output. For instance, in the particular case with $n$ equal output probabilities $p^\text{out}_k{=}{1}{/}{n}$, the conversion rate scales as $\mathcal{C}_n^\text{(passive)}{=}\left(1{/}n\right)^n$, showing the non scalability of passive approaches. On the other hand, an active demultiplexer with time-varying relative output signals $\mathcal{S}^\text{out}(t)$ results in a conversion rate $\mathcal{C}_n^\text{(active)}{=}\frac{1}{T}\int_{T}\left[\prod_{k=1}^{n}\mathcal{S}^\text{out}(t)\right]\text{d}t$, where the integral is taken over the demultiplexing period $T$. In this active case, the conversion rate is typically polynomial in $n$, thus constituting a scalable approach.
 
Solid-state based demultiplexed multi-photon sources have been reported with both passive~\cite{loredo_boson_2017}, and active schemes~\cite{Lenzini:2017aa,wang_high-efficiency_2017,Wang:2018ab}. Thus far, approaches for active demultiplexing have employed either on-chip architectures~\cite{Lenzini:2017aa} with fast reconfigurable speeds ($20$~MHz), but low device throughput transmission; or have combined high-transmission bulk electro-optic modulators (pockels cells)~\cite{wang_high-efficiency_2017,Wang:2018ab}, but requiring very high voltages (${\sim}2000$~V) for each modulator, and with relatively slow reconfigurable speeds ($0.76$~MHz). Here we make use of resonance-enhanced electro-optical modulators (r-EOMs), QUBIG GmbH, that allow for combining high transmission and fast reconfigurability, while requiring low-voltage control (${\sim}5$ V).

Figure~\ref{fig:1}(II) depicts our demultiplexing scheme for preparing a 3-photon source. The system consists of two cascaded high-transmission (95\%) and synchronised r-EOMs. The first r-EOM is driven at one eight of the laser rep. rate, ${\sim}10$~MHz, which combined with a polarising beam-splitter (PBS) distributes 50~ns long time-bins to either output of the PBS alternately. Similarly, a second r-EOM, driven at one sixteenth of the laser rep. rate, ${\sim}5$~MHz, now routes 100~ns long time-bins to either output of the second PBS. Fibered delays of appropriate length are added here to ensure the simultaneous arrival of the three single-photons to the tritter circuit.

\begin{figure}[hbt!]
\centering
\includegraphics[width=.4\textwidth]{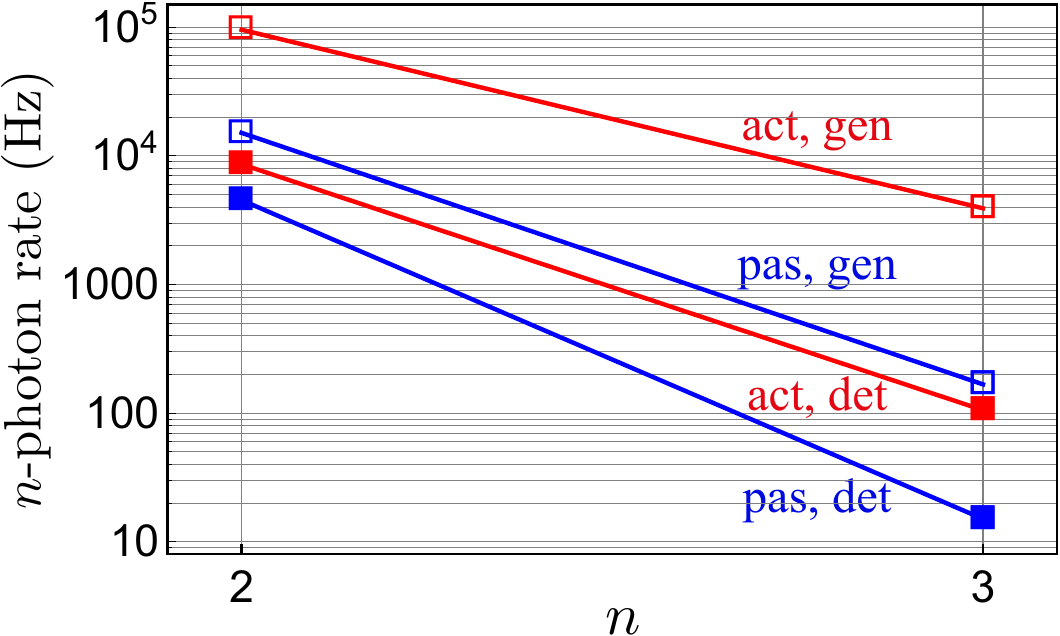}\vspace{-3mm}	
\caption{\textbf{n-photon rates after the demultiplexer.} Generated and detected n-photon rates for the active (red), and passive (blue) demultiplexing schemes.}
\label{fig:3}
\end{figure}

Figure~\ref{fig:2} illustrates the working principle of the demultiplexer. The synchronised operation of both r-EOMs results in the active distribution of consecutive ${\sim}50$~ns long time bins into different spatial outputs, see Fig.~\ref{fig:2}(a). Given the modulation sequences, we can estimate the 3-photon conversion rates $\mathcal{C}_3$ for both the ideal active, and passive scenarios. For the ideal active case, we obtain $\mathcal{C}_3^\text{(active)}{=}1{/}4$; that is, the time-integrated area of the product of all three relative output signals is one fourth of the total demultiplexing area (period) of $T{=}200$~ns. If the demultiplexer is turned off, thus operated as a passive one, the conversion rate now becomes $\mathcal{C}_3^\text{(passive)}{=}1{/}32$---the product of three static transmission probabilities $1{/}4,1{/}4$, and $1{/}2$. The active-to-passive ratio $r_n{=}\mathcal{C}_n^\text{(active)}{/}\mathcal{C}_n^\text{(passive)}$ relates the relative $n$-photon production rates between active and passive schemes, and can be used to assess the demultiplexer's {\it{active efficiency}} $\eta_a{=}({r_n^\text{(exp)}{-}1})/({r_n^\text{(ideal)}{-}1})$---a quantity that equals 1 for an ideal active scheme, and vanishes for a passive one.

Figure~\ref{fig:2}(b) shows our measured output signals for the same demultiplexing period of $T{=}200$~ns, taken using laser light and photodiodes. The non-unity contrast of modulation is due to both imperfect polarisation switching of the r-EOMs, and finite polarisation extinction ratios of the PBSs. Using the same demultiplexing scheme with the single-photon source operated at the increased pump repetition rate of $324$~MHz, we obtain detected, and generated---corrected for detector efficiencies of 0.3---2 and 3-photon rates as shown in Fig.~\ref{fig:3} for both active and passive schemes. The generated (detected) 3-photon rate amounts to $3.8$~kHz ($105$~Hz) for the active scheme. The  active-to-passive ratio is expected to be $r_3^\text{(ideal)}{=}8$, and is measured to reach $r_3^\text{(exp)}{=}6.6$. This corresponds to a 3-photon active efficiency of $\eta_a{=}0.80$, which is here limited by switching contrasts and non-instantaneous modulation raise-, and down-times.

\section{\label{sec:chip} Reconfigurable photonic tritter  chip }

The output of the demultiplexer is connected to a fiber array precisely coupled to the tritter chip inputs. The optical waveguides are fabricated by femtosecond laser writing in a commercial alumino-borosilicate glass substrate (EagleXG, Corning Inc., USA). A Yb:KYW cavity-dumped mode-locked laser oscillator was employed, producing 300~fs duration pulses at 1~MHz repetition rate. In detail, 220~nJ laser pulses were focused 30~$\mu$m below the surface of the glass substrate, by means of a 0.6~N.A. microscope objective, while the sample was translated at the constant speed of 20~mm/s. Such irradiation parameters result in single-mode waveguides at 930~nm wavelength with $\sim$8~$\mu$m $1/e^2$ mode-diameter and $< 1$~dB/cm propagation loss. 

As shown in Fig.~\ref{fig:1}, the tritter is built of three directional couplers (DCs), with nominal reflectivities of $R_1{=}1/2$ (first and last DCs), and $R_2{=}1/3$ (second DC), and one intermediate phase-shifter $\phi$. When the phase shifter is set at $\phi {=} \pi/2$ or $ 3\pi/2$, the theoretical unitary matrix of the circuit corresponds to a symmetric tritter transformation \cite{Zukowski:1997aa,Campos:2000aa,Tichy:2010aa,spagnolo_three-photon_2013}, given by the matrix elements $\mathcal{U}_{jk}^{\textrm{th}}{=}\exp{\left[i 2 \pi (j-1)(k-1) \right]}/\sqrt{3}$, where $j,k{=}1,2,3$ are the corresponding matrix elements indexes. At the input and at the output of the chip the inter-waveguide distance is set to 127~$\mu$m to match the pitch of the fiber arrays. The overall footprint of the waveguide circuit is 25~$\times$~0.25~mm$^2$.

The tuneable phase-shifter $\phi$ is realized according to the method of Refs.~\cite{flamini_thermally_2015, crespi_single-photon_2017}. A 50~nm-thick gold layer is sputtered on the top surface of the chip, and a resistive heater is laser-patterned just above the relevant waveguide (the obtained resistance value is about 60~$\Omega$). By driving the resistor with a suitable voltage, the waveguide is locally heated and increases its refractive index according to the thermo-optic effect, thus producing a finely adjustable phase delay.

\begin{figure*}[hbt!]
\centering
\includegraphics[width=\textwidth]{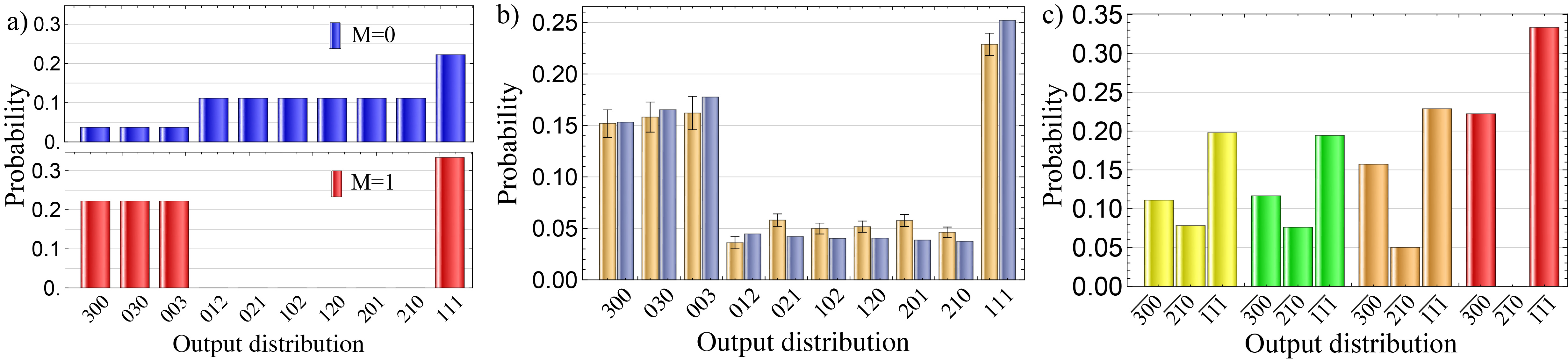}\vspace{-3mm}	
\caption{\textbf{Three photon coalescence.} (a) Theoretical output distributions for an ideal tritter device, and distinguishable (blue) and indistinguishable (red) photons. (b) Experimental output distribution (orange bars), and distribution expected from the modelling of the experiment (light-blue bars). (c) Comparison between output distributions of tritter experiments from~\cite{spring_chip-based_2017} (yellow), ~\cite{spagnolo_three-photon_2013} (green), this work (orange), and the idealised case (red).}
\label{fig:4}
\end{figure*}

We characterize the experimental  matrix $\mathcal{U}^{\textrm{exp}}$ using a continuous-wave laser tuned to the same wavelength as that of the single-photons. We exploit the method described in Ref. ~\cite{Rahimi-Keshari:2013aa} where amplitudes and phases are deduced by measuring the intensity and by monitoring the interferences at the chip outputs when sending the laser directly into one input only or into two inputs simultaneously. For an applied voltage of 3.1 V on the phase shifter, providing the best operation, we obtain a fidelity $\mathcal{F}{=}1{-}\sum_{i{<}j}\sum_{k{<}l} |(\mathcal{V}^{\textrm{th}})_{i,j;k,l}-(\mathcal{V}^{\textrm{exp}})_{i,j;k,l}|/18$ to the ideal tritter $\mathcal{U}^{\textrm{th}}$ of $\mathcal{F} {=}0.960$, where the visibility $\mathcal{V}_{i,j;k,l}$ is given by the ratio $\mathcal{V}_{i,j;k,l}{=}(P^{\textrm{C}}_{i,j;k,l}-P^{\textrm{Q}}_{i,j;k,l})/P^{\textrm{C}}_{i,j;k,l}$, and each of these probabilities are $P^{\textrm{C}}_{i,j;k,l}{=}|\mathcal{U}_{i,k}\mathcal{U}_{j,l}|^2+|\mathcal{U}_{i,l}\mathcal{U}_{j,k}|^2$ and $P^{\textrm{Q}}_{i,j;k,l}{=}|\mathcal{U}_{i,k}\mathcal{U}_{j,l} + \mathcal{U}_{i,l}\mathcal{U}_{j,k}|^2$ (see the supplemental material for a complete description of the chip characterisation). 

\section{\label{sec:results} Three photon coalescence }

Figure~\ref{fig:4}(a) shows the occupation probabilities calculated for an ideal circuit for the ten  three-photon  outputs states $|i,j,k\rangle$  corresponding to $i$ (resp. $j$, $k$) photons in mode 1 (resp. 2, 3). Two cases are considered: fully distinguishable photons with pair wise mean wave-packet overlap $M{=}0$ and three fully indistinguishable photons  with $M{=}1$. For fully indistinguishable photons, the output distribution is composed by the no-collision term $|1,1,1\rangle$, with a probability of 1/3, and by the three-photon bunching terms $|3,0,0\rangle$, $|0,3,0\rangle$ and $|0,0,3\rangle$, with probabilities 2/9 each. In this case, the six possible terms of the output state, having exactly 2 photons in one of the modes and 1 photon in another one, completely vanish. As a result, the eventual detection of such events  indicates the presence of imperfect single-photon indistinguishability. The case of complete distinguishability indeed shows a different distribution with a maximal probability of 1/9 for these states, a reduced probability of 1/27 for the bunching terms and 2/9 for the non-collision term~\cite{Tillmann:2015aa}. 

To experimentally reconstruct the population distribution at the output of the chip, we use a pseudonumber-resolving detection scheme, see module (IV) of Fig.~\ref{fig:1}. Each output is coupled to  two cascaded fibered BSs connected to three silicon APDs. An electronic correlation allows one to reconstruct all the triple photon coincidences. We accumulated a total of 3078 detected triple events during ${\sim}1.7$ hours, collected within a 2 ns  coincidence window. The corresponding reconstruction of the output state is shown in Fig.~\ref{fig:4}(b) in blue bars. There is a strong contribution of the three-photon bunching terms and the non-collision term with an average probability of $\overline{P}_{\{|3,0,0\rangle\}}{=}0.157\pm0.015$ and $P_{|1,1,1\rangle}{=}0.229\pm0.011$, respectively, and an average probability for the terms associated to distinguishable photons $\overline{P}_{\{|2,1,0\rangle\}}{=}0.050\pm0.006$. 

We compare our experimental results with the theoretical distribution  calculated for the measured mean wave-packet overlap and the non-zero $g^{(2)}(0)$. The effect of non-perfect indistinguishability is accounted for following the  model of generalised multiphoton interference by Tillmann and coworkers~\cite{Tillmann:2015aa}. However, such model does not consider the  generation of more than one photon per pulse. To do so, one needs to identify the origin of non perfect  $g^{(2)}(0)$, since different contributions are expected whether the extra photons are identical to the others or not. {Here, we use a neutral exciton under resonant excitation, in which case the non-zero $g^{(2)}(0)$ is due to imperfect suppression of the excitation laser. Inserting a narrow-band spectral filter would further remove the residual laser, leading to almost perfect single-photon purity. The additional photons are thus distinguishable from the QD emission.} We quantify the relative amount of average photon number coming from the QD emission, $\mu_{\textrm{QD}}{=}p_1^{\textrm{QD}}$, and from the scattered laser, $\mu_{\textrm{L}}$, according to the formalism of the probability generating function \cite{Goldschmidt:2013aa} (see supplementary material). The second order autocorrelation function that results from the mixture of the independent photonic distributions of scattered laser and QD single-photons is given by $g^{(2)}(0){=}\chi (2 + \chi)/(1 + \chi)^2$, where $\chi{=}\mu_{\textrm{L}}/\mu_{\textrm{QD}}$. Since we measure a high single-photon purity, we approximate the probability of having a single-photon from the laser by $p_1^{\textrm{L}}{\simeq}( g^{(2)}(0)/2) p_1^{\textrm{QD}}$, neglecting higher order Fock terms from the laser $p_{n{>}1}^{\textrm{L}}\ll p_1^{\textrm{L}}$. 


Following this assumption, we consider that for each  input mode there is a certain probability of having vacuum, a single-photon from the QD, a single-photon from the laser, or most unlikely simultaneous photons from QD and laser. The total sum of these probabilities is normalised such that $p_0+p_1^{QD}+p_1^{L}+p_1^{QD}p_1^{L}{=}1$. 
In this context and considering that $p_0{\gg}p_1^{\textrm{QD}}$, the relevant input tritter states involving three photons are: {$|1_{\textrm{QD}},1_{\textrm{QD}},1_{\textrm{QD}}\rangle$ with probability $(p_1^{\textrm{QD}})^3$, the six combinations of $\{|1_{\textrm{QD}}1_{\textrm{L}},1_{\textrm{QD}},0\rangle\}$ with probability $p_0 (p_1^{\textrm{QD}})^2 p_1^{\textrm{L}}$, and the three combinations of $\{|1_{\textrm{QD}},1_{\textrm{QD}},1_{\textrm{L}}\rangle\}$ with probability ($(p_1^{\textrm{QD}})^2p_1^{\textrm{L}}$)}. 
The total  output distribution is obtained by summing the weighted contributions of all relevant input states and considering the corresponding pair-wise indistinguishabilities between photons generated from the QD, and $M{=}0$ between QD single-photons and laser light. 

Our source single-photon purity is measured to be $g^{(2)}(0){=}0.035{\pm}0.003$. However, when running our experiments with a laser repetition rate of $324$~MHz, we observe a slightly deteriorated effective single-photon purity of $g^{(2)}(0){=}0.071{\pm}0.003$, due to the slow jitter time of our detectors, which we take into account in our modelling, see Supplementary Material. Figure~\ref{fig:4}(b) shows our experimental results, displaying a good agreement between our measurements and simulations.

\begin{table} [t]
\begin{center}
\begin{tabular}{|p{3,2cm}|c|c|c|}
\hline
   & Ref. \cite{spagnolo_three-photon_2013} & Ref. \cite{spring_chip-based_2017} & This work \\ \hline
Source technology & SPDC & SPDC & QD\\ \hline
Integrated source & $\times$ & \checkmark & \checkmark\\ \hline
Integrated Circuit & \checkmark & $\times$ & \checkmark\\ \hline
$g^{(2)}(0)$ & 0.05 & 0.08 & $0.035$\\ \hline
Pair-wise indistinguis. & 0.90 - 0.63 & $0.93$ & 0.90 - 0.88 \\ \hline
3-photon generated rate (Hz)& 20 & $43$ & $ 3800$\\ \hline
3-photon generated rate at tritter output (Hz) & 0.1&  $ 9$ & $ 19$ \\ \hline

\hline\end{tabular}
\end{center}
\caption{Comparative table of the various tritter implementations.}\label{table1}
\end{table}



\begin{table} [t]
\begin{center}
\begin{tabular}{|p{3,5cm}|c|c|} 
\hline
   & This work & Optimised values  \\ \hline
Repetition rate (MHz) & 324 & 1000 \\ \hline 
First lens brightness & 0.16 & 0.65\\ \hline
Fibered brightness & 0.07 & 0.50 \\ \hline
Demultiplexer transmission (per photon) & 0.63 & 0.85\\ \hline
Chip transmission + fiber-array coupling per photon & 0.17 & 0.60\\ \hline
Detection efficiency per photon & 0.30 & $0.9$\\ \hline
Gen. (Det.) 3-photon rate source (Hz)& $3.8{\times}10^3$ (105) & $16{\times} 10^6$ ($12{\times} 10^6$)  \\ \hline
Gen. (Det.) 3-photon rate after chip (Hz) & $ 19$ ($ 0.5$) & $5.5{\times} 10^6$ ($4.0{\times} 10^6$)\\ \hline
Output det. 10-photon rate (Hz) & - &  $40$ \\ \hline

\hline\end{tabular}
\end{center}
\caption{Efficiency budget for the total architecture and foreseen progress in the near future (see details in the text). }\label{table2}
\end{table}

\section{\label{sec:discussion} Discussion \& Conclusion}
Table~\ref{table1} shows various figures-of-merit of sources used in implementations of three-photon interference in a tritter device. Previous experiments used an integrated platform for either the source, or the photonic circuit, but not both simultaneously. The present implementation is the first combining both integrated platforms, already showing a significant  improvement in performance.
 
The brightness of the QD single-photon source, combined with an active demultiplexer provides an increase by at least two orders-of-magnitude of the three-photon generation rate. Moreover, also the tritter output three-photon rate is increased by a factor of ${\sim}2$, and its distribution is closer to the ideal one compared to previous works, see Fig.~\ref{fig:4}(c). These significant advances have been obtained despite non-minimised losses in the present implementation (see Table~\ref{table2}), and using low-efficiency detectors with limited time resolution. The present study thus constitutes a first step on improving the scalability that can be obtained by merging solid-state photon sources and reconfigurable chips.
  
Given the challenges imposed in this first merging of scalable photonic platforms, where not all parts of the system could be optimised at once, we present in Table~\ref{table2} a loss-budget indicating values within reach in the foreseeable future. On the source side, the first lens brightness was  limited by the use of a resonant excitation scheme that removes more than 50\% of the single-photons. Recent studies show that such excitation scheme could be overcome using side excitation \cite{ates_post-selected_2009} or simply removed by taking advantage of Raman assisted excitation \cite{reindl_highly_2019}. We thus anticipate that the first lens brightness using this new excitation scheme could reach the maximal value resulting from the Purcell acceleration and outcoupling efficiency, typically 65\% for the present technology \cite{somaschi_near-optimal_2016}. The fibered brightness was limited here to 7\% due to the use of a microscope objective with relatively small numerical aperture within our collection setup (N.A.{=}0.45). Inserting a high numerical aperture collection lens inside the cryostat has been shown to solve this problem \cite{wang_high-efficiency_2017}. Moreover, recent technological progress on the glass chip technology shows that its throughput transmission could reach much higher values, by adopting proper post-processing with thermal-annealing after the waveguide inscription \cite{arriola_low_2013}. Considering all these possible improvements, we anticipate that one could soon reach on-chip ten-photon manipulation at rates as high as ${\sim}40$~Hz.

 \vspace{0.5cm}
\begin{acknowledgments}

{\bf Acknowledgments:} The authors acknowledge financial support by the QuantERA ERA-NET Cofund in Quantum Technologies, project HIPHOP.  P.S. acknowledges further support by the ERC Starting Grant No.~277885 QD-CQED, the French Agence Nationale pour la Recherche (grant ANR SPIQE and USSEPP), the French RENATECH network, a public grant overseen by the French National Research Agency (ANR) as part of the ``"Investissements d'Avenir"• programme (Labex NanoSaclay, reference: ANR-10-LABX-0035). C.A. and J.C.L. acknowledge support from Marie Sk\l{}odowska-Curie Individual Fellowships SQUAPH and SMUPHOS, respectively. R.O. acknowledges support from the ERC Advanced Grant CAPABLE (grant no. 742745).
\end{acknowledgments}

\bibliography{tritter.bib}

\end{document}